\documentclass[conference]{IEEEtran}
\IEEEoverridecommandlockouts

\usepackage{amsmath,amssymb,amsfonts}
\usepackage{graphicx}
\usepackage{textcomp}
\usepackage{xcolor}
\usepackage{adjustbox}
\usepackage{hyperref}
\usepackage{cleveref}

\DeclareMathOperator*{\argmax}{argmax}

\usepackage[backend=biber,doi=false,style=ieee,natbib=true]{biblatex} 

\usepackage{longtable,tabularx,booktabs}
\usepackage[flushleft]{threeparttable}
\usepackage{colortbl}
\usepackage{mdframed}
\mdfdefinestyle{MyFrame}{%
    linecolor=black,
    outerlinewidth=2pt,
    roundcorner=5pt,
    innertopmargin=\baselineskip,
    innerbottommargin=\baselineskip,
    innerrightmargin=15pt,
    innerleftmargin=15pt,
    backgroundcolor=white
}

\def\BibTeX{{\rm B\kern-.05em{\sc i\kern-.025em b}\kern-.08em
    T\kern-.1667em\lower.7ex\hbox{E}\kern-.125emX}}

\usepackage{algorithmicx}
\usepackage{algorithm}
\usepackage{algpseudocode}
\algtext*{EndLoop} 
\algtext*{EndIf}
\algtext*{EndFunction}

\begin{document}

\title{On Technique Identification and Threat-Actor Attribution using LLMs and Embedding Models}

\author{\IEEEauthorblockN{Kyla Guru}
\IEEEauthorblockA{\textit{Computer Science} \\
\textit{Stanford University}\\
Stanford, CA \\
kylaguru@stanford.edu}
\and
\IEEEauthorblockN{Robert J. Moss}
\IEEEauthorblockA{\textit{Computer Science} \\
\textit{Stanford University}\\
Stanford, CA \\
mossr@cs.stanford.edu}
\and
\IEEEauthorblockN{Mykel J. Kochenderfer}
\IEEEauthorblockA{\textit{Aeronautics and Astronautics}}
\textit{Stanford University}\\
Stanford, CA \\
mykel@stanford.edu}

\maketitle

\begin{abstract}
Attribution of cyber-attacks remains a complex but critical challenge for defenders. Currently, manual extraction of behavioral indicators from dense forensic documentation causes significant attribution delays, especially following major incidents at the international scale. This research evaluates large language models (LLMs) for cyber-attack attribution based on behavioral indicators extracted from forensic documentation. We test OpenAI's GPT-4 and \textit{text-embedding-3-large} for identifying threat actors' tactics, techniques, and procedures (TTPs) by comparing LLM-generated TTPs against human-generated data from MITRE ATT\&CK Groups. Our framework then identifies TTPs from text using vector embedding search and builds profiles to attribute new attacks for a machine learning model to learn. Key contributions include: (1) assessing off-the-shelf LLMs for TTP extraction and attribution, and (2) developing an end-to-end pipeline from raw CTI documents to threat-actor prediction. This research finds that standard LLMs generate TTP datasets with noise, resulting in a low similarity to human-generated datasets. However, the TTPs generated are similar in frequency to those within the existing MITRE datasets. Additionally, although these TTPs are different than human-generated datasets, our work demonstrates that they still prove useful for training a model that performs above baseline on attribution. Project code and files are contained here: \url{https://github.com/kylag/ttp_attribution}. 
\end{abstract}


\section{Introduction}
Technical attribution involves identifying the originating group behind a cyber-attack. This attribution is a complex, but mission-critical task for defenders. For countries with less access to cybersecurity capabilities, delays in attribution can lead to serious diplomatic and geopolitical ramifications. For instance, after the 2021 Microsoft Exchange Server Campaign, NATO and EU member states failed to issue a timely condemnation of China's use of contract hackers due to delays in independent attribution \cite{soesanto2021divided}. 

Attribution of cyber-attacks by an all-source analyst is broken down into two core steps: (1) parsing unstructured forensic reports to identify threat actor's tactics, techniques, and procedures (TTPs) of the adversary (i.e., TTP identification), and (2) using TTPs, among other indicators, to predict a specific threat group for attribution. Prior work demonstrated the effectiveness of AI-based approaches to the attribution pipeline \cite{egyptian,wang2021explainable,abdi2023automatically}. Specifically, natural language processing (NLP) techniques have proven successful for \textit{step 1} of extracting information about APT behavior from post-incident documentation \cite{perry2019no,DeepLearningForThreats,irshad2023cyber}. Additionally, more classical machine learning (ML) techniques have been applied to \textit{step 2}, suggesting that models have the ability to learn patterns of behavior and make predictions given new sets of inputs (e.g., documentation that has not been used in training the model) \cite{egyptian}. In many ways, cyber-attack attribution can be represented as a multi-class prediction problem with multiple \textit{advanced persistent threat} (APT) groups originating from different countries \cite{zhang2024attribution}.

 Recent advances suggest that large language models (LLMs) excel at understanding and generating human-like text, making them well-positioned for tasks like translating language, summarizing text, and question-answering \cite{matarazzo2025survey}. Given the unstructured nature of threat intelligence, we investigate how well an LLM performs at tasks related to attribution, particularly observing the impacts of an LLM's large training space on low-data, high-stakes decisions.



In this study, we compare LLM generated TTPs to TTPs published on MITRE's threat intelligence platform, using MITRE's dataset as a proxy for human-generated TTP sets that we consider to be the``ground truth" \cite{mitre2024groups}. Then, we use LLM-generated TTPs to train a machine learning model and build an end-to-end pipeline that takes in raw documentation as input, and produces a TTP-based attribution prediction as the output. We evaluate the model's attribution performance by examining the rank position of the correct threat actor in the model's prediction list, where a rank of 1 indicates the model correctly identified the threat actor as its top choice.



This work reveals the strengths and weaknesses of applying LLMs for cybersecurity attribution, providing critical guidance as off-the-shelf LLM tools increasingly enter high-stakes security decision-making. This paper's major contributions include: (1) assessing off-the-shelf LLMs and vector embedding search for TTP extraction and attribution; and (2) developing an end-to-end pipeline from raw CTI documents to threat-actor prediction. Our findings show that while LLM-generated TTPs differ significantly from human-annotated datasets, they maintain frequency patterns aligned with MITRE ATT\&CK and enable above-baseline attribution performance—particularly for threat actors with distinctive TTP profiles.


\section{Background and Related Works}
While prior approaches have focused on conducting attribution by learning patterns within binary source code or malware through traditional AI-based approaches such as pattern recognition, clustering algorithms, or feed-forward neural networks \cite{quiring2019misleading,kida2023nation}, limited prior research has studied the potential of LLMs to be used for the extraction of TTPs from heterogeneous CTI reports for attribution. 
In the literature on attribution, cyber-attackers that are well-resourced and sophisticated that typically target high-value organizations are referred to as \textit{advanced persistent threat} (APT) groups \cite{crowdstrike2023apt}. As a result of increased monitoring and sharing around the activity of these groups, the TTPs of APT groups have become more widely researched and publicly disclosed. For example, CrowdStrike currently tracks over 150 APT groups all over the world, including nation-states, eCriminals, and hacktivists \cite{crowdstrike2023apt}.  Most APT groups are funded or operated by nation-states, making these groups extremely skilled at advanced obfuscation and deceptive techniques \cite{xu2024comprehensive}.

\subsection{TTP Identification}
The standard practice for extraction is based on a manual process of obtaining training data for a model. More recent studies have extracted vector representations of threat actor behavior from unstructured CTI reports using various natural language processing (NLP) techniques, and then used this data to train a model for the attribution task  \cite{Jo2022,perry2019no,DeepLearningForThreats,egyptian}. Studies using LLMs have attempted to identify TTPs based on CTI documentation using GPT-3.5 \cite{AutomatedMapping}, and have benchmarked various LLMs on their ability to attribute threat actors based on redacted CTI documents and prompt engineering. However, these studies have not yet tested an end-to-end pipeline that takes in documentation and returns an attribution prediction using LLM-based TTP techniques.

\subsection{APT Attribution}
The pipeline of technical attribution also tends to be a highly manual process that begins right after an incident occurs and may last for months, if not years, beyond the time of the attack \cite{AttributionLong}.
Researchers have looked at ML-based attribution using low-level \textit{indicators of compromise} (IOCs) such as IP addresses and hash values, or using code authorship techniques with malicious binary files \cite{noor2023machine,alrabaee2016feasibility,gray2021identifying}. However, the literature describes that these types of low-level indicators can be altered easily or masked by anonymous proxy services like Tor \cite{bianco2013pyramid}.

There have also been research studies on using dynamic analysis of malware in a sandbox to observe program behavior, and using this behavior for technical attribution. For instance, several studies have extracted words such as system calls from Cuckoo Sandbox reports \cite{david2015deepsign,jamalpur2018dynamic,chen2017automated,rosenberg1970deepapt}, and then used these extracted ``features'' to train an attribution classifier to classify threat groups from new report samples \cite{shiva2016windows, APTMalware2023}.

However, recent literature argues that code authorship attribution is also becoming less helpful for cyber attack attribution, as it is susceptible to adversarial attacks that can cause up to a 99\% reduction in accuracy \cite{quiring2019misleading}. APT groups also now enable false attribution through shared code bases and by switching to languages like Go, Rust, and Dlang that lack accurate attribution mechanisms due to their newer compilation technologies and limited compiler fingerprinting capabilities compared to traditional languages.\cite{kida2023nation}. 


The literature argues that a group's behavioral indicators, otherwise known as \textit{tactics, techniques, and procedures} (TTPs), are more difficult to alter or obfuscate \cite{bianco2013pyramid}. While other indicators such as code styles, IP addresses, and target sector might be more ephemeral and easy for a threat actor to alter, changing one's TTPs increases cost of an operation. The adversary must now make the time-consuming effort to learn new behaviors, change their tools and techniques, reinvent themselves from scratch, or give up entirely \cite{irshad2023cyber}.


More recently, \cite{noor2023machine} studied various ML approaches using TTPs to conduct attribution. This research was limited to manual extraction of TTPs, suggesting that other techniques such as LLM-based approaches may prove useful for processing large volumes of complex information. This research contributes a framework for both processing unstructured textual evidence, and using these derived technical indicators to make informed decisions. 




\section{Methodology}



The methodology is split into two sub-tasks: (1) TTP identification and (2) threat-actor attribution. TTP identification was conducted in two ways: the first approach prompts a pre-trained LLM for TTPs represented in a document, and the second approach uses vector-embedding search to find similarities between content in documents and TTP descriptions.

\subsection{Step 1. TTP Identification}
TTP identification requires an input set of documents for each threat actor, and results in an output of TTPs representing each threat actor. The formalized notation is as follows. For threat actors $t$ defined as $t \in \{1,\ldots,m\}$, we use $\mathbf{x}_\text{doc}^{(t)}$ to denote the set of reference documents that are associated with one threat actor. Each individual document $i$ for the threat actor $t$ is represented as $\mathbf{x}_\text{doc}^{(t,i)}$. For each threat actor $t$, we denote $\mathbf{y}_\text{ttp}^{(t)}$ to be the human-generated TTPs.  
Across all threat actors, the set of documents representing every threat actor is represented as $\mathcal{X}_\text{doc}^{(t)}$, and the output, or the set of all human-generated TTPs representing every threat actor is represented as $\mathcal{Y}_\text{ttp}^{(t)}$: 
\begin{gather}
\mathcal{X}_\text{doc}^{(t)} = \big\{ \mathbf{x}_\text{doc}^{(t,i)} \mid i \in 1,\ldots,n_t \big\} \\
\mathcal{Y}_\text{ttp}^{(t)} = \big\{ \mathbf{y}_\text{ttp}^{(t,i)} \mid i \in 1,\ldots,n_t \big\}
\end{gather}

For each document $i$, $\mathcal{X}_\text{doc}^{(t)}$, $\mathbf{y}_\text{ttp}^{(t,i)}$ represents the human-generated TTPs. For each threat actor, documents and their associated TTPs are combined to create dataset $\mathcal{D}_\text{ttp}^{(t)}$, where the full dataset across all threat actors is defined as $\mathcal{D}_\text{ttp}$:
\begin{gather}
    \mathcal{D}_\text{ttp}^{(t)} = \Big\{ \big(\mathbf{x}_\text{doc}^{(t,i)}, \mathbf{y}_\text{ttp}^{(t,i)}\big) \mid \mathbf{x}_\text{doc}^{(t,i)} \in \mathcal{X}_\text{doc}^{(t)},\, \mathbf{y}_\text{ttp}^{(t,i)} \in \mathcal{Y}_\text{ttp}^{(t)} \Big\} \\
    \mathcal{D}_\text{ttp} = \big\{ \mathcal{D}_\text{ttp}^{(t)} \mid t \in 1,\ldots, m \big\}
\end{gather}
The task of predicting TTPs given a document is performed using a TTP prediction model $\mathcal{M}_\text{ttp}$:
\begin{gather}
\tilde{\mathbf{y}}_\text{ttp}^{(t,i)} = \mathcal{M}_\text{ttp}\big(\mathbf{x}_\text{doc}^{(t,i)}; \theta\big)
\end{gather}
Two separate modeling approaches for TTP identification were studied in this work:
\begin{enumerate}
    \item Prompting a pre-trained LLM with parameters $\theta_\text{LLM}$ (e.g., prompt shown in \cref{fig:gpt_prompt}).
    \item Using vector embedding search with model parameters $\theta_\text{ve}$ to identify TTPs that are close to one another in the embedding space (shown in \cref{alg:ve_identification}). 
\end{enumerate}
When prompting an LLM for TTPs, the model is treated as a black-box and simply queried with a prompt included in $\theta_\text{LLM}$:
\begin{equation}
    \tilde{\mathbf{y}}_\text{ttp}^{(t,i)} = \mathcal{M}_\text{ttp}(\mathbf{x}_\text{doc}^{(t,i)}; \theta_\text{LLM}) \label{eq:llm}
\end{equation}
where \cref{fig:gpt_prompt} shows an example prompt that generates a parse-able list of identified TTPs from a document.

Alternatively, TTP identification can be done using vector embedding (VE) search, shown in \cref{alg:ve_identification}. The VE search begins by splitting each document into chunks of size $k$ lines where one chunk is labeled as $\ell_k$. An embedding is generated for each chunk $\ell_k$, denoted as $\tilde{\mathbf{z}} = \mathcal{E}(\ell_k; \theta_\text{ve}) \text{ where } \ell_k \in \mathbf{x}_\text{doc}^{(t,i)}$ and $\mathcal{E}$ is the embedding model.

\begin{algorithm}[!t]
    \caption{TTP identification with vector embedding search.} 
    \label{alg:ve_identification}
    \begin{algorithmic}[1]
    \Require $\mathbf{x}_\text{doc}$: CTI document.
    \Require $\theta_\text{ve}$: Parameters of the vector embedding model.
    \Require $\mathcal{T}$: Set of TTP numbers and definitions.
    \Require $k$: Number of lines in the document window.
    \Function{TTP-Identification}{$\mathbf{x}_\text{doc}, \theta_\text{ve}, \mathcal{T}, k$} 
        \State $\tilde{\mathcal{T}}_\mathcal{E} = \{ \mathcal{E}(y; \theta_\text{ve}) \mid y \in \mathcal{T} \}$ \Comment{Vector embedded TTPs} \label{line:start}
        \State $\tilde{\mathbf{y}}_\text{ttp} = \emptyset$
        \For{$\ell_k \in \mathbf{x}_\text{doc}$}
            \State $\tilde{\mathbf{z}} = \mathcal{E}(\ell_k; \theta_\text{ve})$ \Comment{Vector embedding for lines $\ell_k$}
            \State $\tilde{\mathbf{t}} = \argmax_{\tilde{\mathbf{t}} \in \tilde{\mathcal{T}}_\mathcal{E}} \frac{\tilde{\mathbf{z}} \cdot \tilde{\mathbf{t}}}{\lVert \tilde{\mathbf{z}} \rVert \lVert \tilde{\mathbf{t}} \rVert}$ \Comment{Embedding similarity}
            \State $\tilde{y}_\text{ttp} = \mathcal{C}(\tilde{\mathbf{t}})$ \Comment{Decode embedding to TTP}
            \State Append $\tilde{y}_\text{ttp}$ to $\tilde{\mathbf{y}}_\text{ttp}$
        \EndFor \label{line:end}
    \State \Return $\tilde{\mathbf{y}}_\text{ttp}$
    \EndFunction
    \end{algorithmic}
\end{algorithm}

A dataset $\mathcal{T}$ of TTPs $y$ and their corresponding definitions is used for the vector embedding search method. We precompute the embeddings that represent each TTP and their description to generate the list of all TTPs and their corresponding vectorized embedding; denoted as $\tilde{\mathcal{T}}_\mathcal{E} = \{ \mathcal{E}(y; \theta_\text{ve}) \mid y \in \mathcal{T} \}$.

To determine the TTP representing the chunk of lines $\ell_k$ from the document, the cosine similarity is taken between the vector embedding $\tilde{\mathbf{z}}$ and each TTP $y$ in the $\tilde{\mathcal{T}}_\mathcal{E}$ dataset. The TTP $y$ with the highest cosine similarity is then tagged as the TTP representing the chunk of lines:
\begin{gather}
\tilde{\mathbf{t}} = \argmax_{\tilde{\mathbf{t}} \in \tilde{\mathcal{T}}_\mathcal{E}} \frac{\tilde{\mathbf{z}} \cdot \tilde{\mathbf{t}}}{\lVert \tilde{\mathbf{z}} \rVert \lVert \tilde{\mathbf{t}} \rVert}
\end{gather}
then we decode to get the TTP using $\tilde{y}_\text{ttp} = \mathcal{C}(\tilde{\mathbf{t}})$.

For either the LLM-prompted or VE approach, the predicted TTPs identified across documents for a threat actor are then processed to create a single unique set of TTPs representing the behavioral indicators of that threat actor group:
\begin{gather}
\tilde{\mathcal{Y}}_\text{ttp}^{(t)} = \big\{ \tilde{\mathbf{y}}_\text{ttp}^{(t,i)} \mid i \in 1, \ldots, n_t \big\}
\end{gather}
The count associated with a TTP is incremented each time it appears across documents.

\begin{algorithm}[!t]
    \caption{Threat actor attribution given a set of TTPs.} 
    \label{alg:attribution}
    \begin{algorithmic}[1]
    \Require $\tilde{\mathbf{y}}_\text{ttp}$: Predicted TTP list (from either LLM or VE).
    \Require $\tilde{Y}$: Pre-trained TTP/APT weights matrix (from VE).
    \Function{ThreatActorAttributon}{$\tilde{\mathbf{y}}_\text{ttp}, \tilde{Y}$} 
        \State Normalize counts in $\tilde{\mathbf{y}}_\text{ttp}$ to get $\bar{\mathbf{y}}$
        \State $\mathbf{p}_\text{attr} = \bar{\mathbf{y}}^\top \tilde{Y}$
    \State \Return top-$r$ ranked threat actors from $\mathbf{p}_\text{attr}$
    \EndFunction
    \end{algorithmic}
\end{algorithm}

\subsection{Step 2. Attribution Using Identified TTPs}
Threat actor attribution (\cref{alg:attribution}) outputs a ranked list of potential threat groups based on identified TTPs (agnostic to the type of model that generated the TTP predictions $\tilde{\mathbf{y}}_\text{ttp}$).
Before threat actors are attributed given new data, a normalized weights matrix $\tilde{Y}$ is trained using the same VE search \cref{alg:ve_identification}, but now over a \textit{training dataset} of documents.
After TTP identification is completed across all documents in the training dataset, the weights matrix $\tilde{Y}$ is computed by normalizing the counts of each TTP associated to each threat actor. The columns of $\tilde{Y}$ are the normalized counts of each TTP, where the rows refers to a threat actor $t$, such that each row sums to $1$.
Treating $\tilde{\mathbf{y}}_\text{ttp}$ as a set of counts for each TTP predicted to be in the input document, we then normalize the counts, labeled as $\mathbf{\bar{y}}$, and multiply with the weights matrix:
\begin{gather}
    \mathbf{p}_\text{attr} = \mathbf{\bar{y}}^\top \tilde{Y}
\end{gather}
where $\mathbf{p}_\text{attr}$ can be understood as a probability distribution of attribution over all threat actors for the new document.
The top-$r$ attributed threat actors are ranked based on their value in $\mathbf{p}_\text{attr}$ and returned from \cref{alg:attribution} as the prediction.

\section{Experimentation}
To test the LLM-based TTP identification procedure in \cref{eq:llm}, we use OpenAI's GPT-4 model, as it represents the state of the art model at the time of writing. We do not use any fine-tuned models since a preliminary benchmark of off-the-shelf LLM performance on attribution-based tasks did not exist prior to this study. For the vector embedding search approach, we use OpenAI's \textit{text-embedding-3-large} pre-trained embeddings model; an embedding model that has been trained on content across websites, blogs, news articles, and other publicly accessible web pages. OpenAI's vector embedding model demonstrates strong performance among commercially available embedding models, making it a suitable choice for our testing \cite{muennighoff-etal-2023-mteb}. 

\subsection{Step 1. TTP Identification}

\begin{figure}[t]
    \begin{mdframed}[style=MyFrame]
    \textbf{GPT Prompt:} You are a cybersecurity analyst. Please read through this attached document and identify all the most important TTPs of the \textbf{[APT name]} threat actor that are described in this document that map to the MITRE ATT\&CK framework. Output specific technique ID numbers and technique ID names identified, including if there is a sub-technique that the threat actor uses. If the threat actor appears to attack the critical infrastructure or Industrial Control System (ICS) sectors, please focus on ICS-related techniques in the output as well (e.g., TTPs beginning with T0). \\
    
    Please output all this information in the form of a comma-separated list. Limit it to the top TTPs that you find are the most important that describe this threat actor's behavior, separated in a new line. Limit TTPs outside of the MITRE list. Example:\\
    {\footnotesize\texttt{[\textquotesingle{}T1083\textquotesingle{},\textquotesingle{}File and Directory Discovery\textquotesingle{}],}}\\
    {\footnotesize\texttt{[\textquotesingle{}T1588\textquotesingle{},\textquotesingle{}.002\textquotesingle{},\textquotesingle{}Obtain Capabilities: Tool\textquotesingle{}]}}
    \end{mdframed} 
    \caption{GPT prompt used to predict TTPs from an attached document.}
    \label{fig:gpt_prompt}
\end{figure}

To generate the dataset of documents across various threat actors, $\mathcal{X}_\text{doc}^{(t)}$, the MITRE ATT\&CK Groups page is used, and references for each threat actor group on MITRE are processed for the task of TTP identification \cite{mitre2024groups}. On any given group's page, there are anywhere between 2--55 references (raw post-incident reports) cited by analysts, followed by the TTPs identified from each report. Because of this association between the cited references and extracted TTPs from those references, we can use the MITRE dataset as a proxy to represent a human-generated TTP dataset in this study.

We analyze the two different approaches for the sub-task of TTP identification: prompting an LLM for a response ($\theta_\text{LLM})$ and vector embedding search ($\theta_\text{ve}$). The GPT prompt used in this work is shown in \cref{fig:gpt_prompt}.

When generating embeddings for TTPs and their definitions for the vector embedding search approach, the MITRE ATT\&CK Techniques are also used with their corresponding definitions listed on MITRE's pages \cite{mitre2023techniques}. Another follow-on approach known as \textit{hypothetical document embeddings} (HyDE) \cite{gao2022precise}, was also used to augment the embeddings of TTP definitions, or $\tilde{\mathcal{T}}_\mathcal{E}$, which is described in the results below.

Using \textit{text-embedding-3-large}, embeddings are also calculated for each $k = 3$ lines of the document, resulting in $\tilde{\mathbf{z}} = \mathcal{E}(\ell_k; \theta_\text{ve}) \text{ where } \ell_k \in \mathbf{x}_\text{doc}^{(t,i)}$. From experimentation, it was found that the accuracy metrics performed better without a sliding window approach over a single document.

After generating the predicted TTPs $\tilde{\mathcal{Y}}_\text{ttp}^{(t)}$ using both approaches ($\theta_\text{LLM}$ and $\theta_\text{ve}$), the results are compared with MITRE's human-generated TTP set associated with each threat actor group ${\mathcal{Y}}_\text{ttp}^{(t)}$ using \textit{Jaccard similarity}. This metric quantifies the overlap between predicted and ground truth TTP sets, yielding values in the range $[0,1]$, where $1$ indicates perfect alignment and $0$ signifies no overlap.
\begin{gather}
J\Big(\tilde{\mathcal{Y}}_\text{ttp}^{(t)},\mathcal{Y}_\text{ttp}^{(t)}\Big) = \frac{\big|\tilde{\mathcal{Y}}_\text{ttp}^{(t)} \cup \mathcal{Y}_\text{ttp}^{(t)}\big|}{\big|\tilde{\mathcal{Y}}_\text{ttp}^{(t)} \cap \mathcal{Y}_\text{ttp}^{(t)}\big|}
\label{eq:jaccard}
\end{gather}

Jaccard similarity is used for capturing similarities and differences between two distinct datasets, and is calculated here for both technique and sub-technique IDs \cite{manning2008introduction}. The technique ID is typically represented by ``T", followed by a four-digit number, and if there is a sub-technique number associated, the there may also be a sub-technique ID number associated with the technique. This may appear like:
\begin{equation*}
\footnotesize 
\texttt{[\textquotesingle{}T1087\textquotesingle{},\textquotesingle{}.001\textquotesingle{},\textquotesingle{}Account Discovery: Local Account\textquotesingle{}]}
\end{equation*}
where the technique ID is ``T1087," and this specific sub-technique mentioned is sub-technique ``T1087.001" \cite{mitre_t1087_001}.

Similarly for VE-based identification, Jaccard similarity is calculated between the VE-generated TTPs at the technique level, and for the $29$ more well-reported threat actors. \footnote{The most well-reported threat actors were the actors that had 10 or more references related to their behavior on the MITRE webpage, allowing for a more holistic view of threat activity.} Thus, the comparison across the two approaches occurs over $29$ threat actors at the broader technique level.

Specific to the LLM approach, we calculate other metrics including the percentage of TTPs in the MITRE dataset, but not in the GPT-generated dataset (labeled as ``In MITRE, not GPT''). This percentage represents the data that GPT-4 may have omitted when analyzing documentation. We also calculate the percentage of TTPs that are labeled as ``In GPT, not MITRE'' to capture the ``noise" that the LLM generated in its TTP output. The final metric we analyze is the \textit{set difference}, which is the proportion of items unique to one set relative to the total number of items in that set.

\subsection{Step 2. Attribution}

The MITRE ATT\&CK dataset was divided into training, validation, and test sets using multiple 70\% train, 20\% validation, and 10\% test splits. This approach is used to develop a model that is invariant to different samples of training and validation data. To complete the set of $727$ documents, additional data containing additional open-sourced reports on $12$ of the $29$ threat actors is added to the existing documentation gathered from MITRE \cite{ThaiCert,Github}.


\subsubsection{Model Training}

Across $10$ training splits, $10$ separate weight matrices are constructed as a part of the standard $k$-fold cross-validation approach \cite{k-fold}. The performance of the validation split is compared across the $10$ weight matrices and the best-performing matrix for prediction, $\tilde{Y}$, is selected.

The vector embedding search approach is used for all documents in the training dataset to generate $\tilde{\mathcal{Y}}_\text{ttp}$. After generating all the counts of each identified TTP for a threat actor, the count of a single TTP is divided by the total count of all TTPs generated for that single threat actor to create a normalized weights matrix, $\tilde{Y}$.


This normalized vector of values represents the $P(\tilde{y}_\text{ttp}  \mid  t)$ for each predicted TTP $\tilde{y}_\text{ttp}$, for any given threat actor $t$. But we are interested in $P(t \mid \tilde{y}_\text{tt})$, so we use Bayes' theorem:
\begin{equation}
P(t \mid \tilde{y}_\text{ttp}) = \frac{P(\tilde{y}_\text{ttp} \mid t) P(t)}{P(\tilde{y}_\text{ttp})}
\end{equation}
given an expert prior $P(t)$.
We experiment using a uniform prior over threat actors and an `expert prior' which was fit to occurrences in the training data. Note that when comparing the relative probabilities given the same evidence, the denominator in Bayes' theorem $P(\tilde{y}_\text{ttp})$ becomes irrelevant. For example, when comparing \( P(t_1 \mid \tilde{y}_\text{ttp}) \) and \( P(t_2 \mid \tilde{y}_\text{ttp}) \) for the same TTPs, both are divided by the same \( P(\tilde{y}_\text{ttp}) \). 
This results in a simplified comparison of the products of the likelihood and prior probabilities by dropping the denominator. We can use an unnormalized probability when comparing across $t_1, t_2, etc$ because we are fixing the condition of $y_{ttp}$:
\begin{equation}
    P(t \mid \tilde{y}_\text{ttp}) \propto P(\tilde{y}_\text{ttp} \mid t)P(t)
\end{equation}
This method of generating \( P(\tilde{y}_\text{ttp} \mid t) \) for each TTP provides a nuanced understanding of the relative probabilities that specific TTPs are associated with a given threat actor, compared to the binary association from the existing MITRE dataset.

The result of model training are $10$ weights matrices based on the different training splits---in each weights matrix, the rows represent threat actor groups, the columns represent MITRE TTPs, and the value contained within each position represents the $P(t  \mid  \tilde{y}_\text{ttp})$.

\subsubsection{Selecting Weight Matrix for Prediction}

To select the weights that will be used for the prediction, the $20\%$ validation document set associated with each train split is used to find the ``best-performing" weights generated by each train split. To determine the ``best-performing" matrix of weights, the documents within the $20\%$ validation set are processed by using vector embedding search to identify TTPs on documentation not included in the training.



In this case, ``success" is defined as a higher ranking (lower true number), or the model ranking the correct threat actor attributed to a document in rank $\#1$. For example, if ``Lazarus Group" was actually attributed to an incident, the model ranking this group as $\#1$ would be successful, and $\#29$ would be unsuccessful. The weights matrix that generates the highest average rank is considered for the $10\%$ test group.

\subsubsection{Using Weight Matrix for Prediction}
To test the performance of the selected best matrix for the task of prediction, the $10\%$ test set is used to find the overall average performance, or the average ranking, of the model. This approach includes taking each document within the test set and generating the TTP matrix for the document using the vector embeddings approach, running \cref{alg:attribution} using the best weight matrix from the validation phase, and generating a ranked list of predicted threat actors. Using this result, the threat actors suspected to be behind the incident are ranked from $1$ to $29$, where the threat actor with the highest predicted association receives rank 1.

\begin{figure*}[t]
    \centering
    \includegraphics[width=0.8\linewidth]{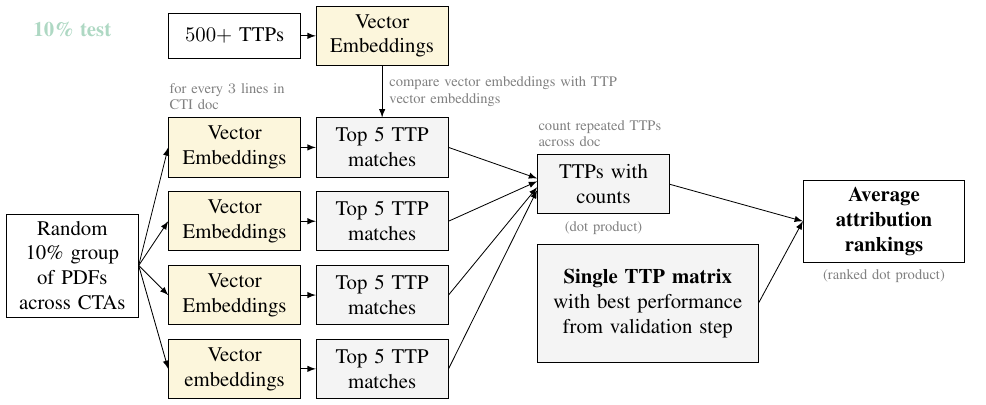}
    \caption{Testing split using unseen 10\% of total documentation.}
    \label{fig:test-split}
\end{figure*}

For all documents within the $10\%$ test set, the position of the correctly attributed threat actor among the $29$ ranked threat actors is found. This is used to compute the average ranking across all the documents in the test set, which represents the final performance of the model.

\section{Results}

\subsubsection{Jaccard Similarity Analysis}

The Jaccard similarity for the TTPs generated by the VE search approach was greater than those generated by the LLM approach. The average Jaccard similarity for the overall technique IDs using GPT-4 is $0.39$, or $39 \pm 12\%$. This means that GPT produces a list of TTPs that represent a single threat actor that has, on average, $39\%$ similarity to the list of TTPs generated by manual analysis, with a high amount of variation across similarity scores assigned to threat actors. 

In the case of using VE search to identify TTPs, the average Jaccard similarity to the human-generated dataset was $18 \pm 7.39\%$ for general technique IDs. On average, the VE approach produced a list of TTPs for any given threat actor that was $18\%$ similar to the human-generated list of TTPs by MITRE.

Overall, the similarity score remains low using for both the GPT-generated TTPs and for TTPs generated through the vector embedding search. This is likely due to the large denominator that represents the union of the two datasets, which may result in a smaller overall numerical output from the Jaccard similarity equation. 



\subsubsection{Set Differences}

Across $140$ threat actors, the data showed that the average percentage of TTPs that were present ``In MITRE, not GPT'' was about $41 \pm 23\%$. While the average difference in coverage between the datasets is $41\%$, the specific difference varies by about $23\%$ points from this average in the dataset.  

For the VE search approach, the average percentage of TTPs present in the MITRE set and not present in the VE TTPs is $42 \pm 11\%$. On average, the TTPs in the vector embeddings set cover over half of the TTPs from the MITRE set, but miss about $42\%$ of the human-generated TTPs.  

On the contrary, the data showed that across $140$ threat actors, the percentage of TTPs that were present in the ``In GPT, not MITRE'' was about $66\%$ in the general case, and about $76\%$ with the added sub-technique. This indicates that the GPT dataset is noisy—without constraints, the model generates as many TTPs as it can find within the provided references, and it remains a black-box whether these TTPs truly come from the documentation, or from GPT's prior knowledge of the incident at large.

For the VE search approach, the data showed that the average percentage of TTPs that were present in the VE dataset and not in the MITRE set was $77 \pm 8.50\%$. This indicates that without fine-tuning, the VE search approach also produces a significant amount of additional TTPs that may not have been extracted from the documentation and represented in the original MITRE dataset. 


\subsection{Attribution Using VE Search: Average Model Ranking}

In the context of attributing the threat actor to a specific set of TTPs, the baseline performance of random guessing can be modeled as a uniform distribution over the integers $1$ to $29$. The mean of a discrete, uniform distribution from $1$ to $29$ is $15$ with a standard deviation of $8.37$.

To assess the performance of attribution model $\tilde{Y}$, the average placement of the correct threat actor among the $29$ total threat actors is considered. The average ranking, or performance of the model, is also calculated with changing parameters such as the expert prior, or the analyst's input of the prevalence of a specific threat actor at the time of prediction.

Shown in \cref{tab:model_performance}, the model achieves better-than-baseline attribution accuracy while the expert prior is uniform across threat actors, which improves performance by approximately three ranks, resulting in an average ranking of $10.96$, or $10.68$. 


Using the expert prior, we see improvements in performance by approximately three more ranks. The average ranking of the best performing model becomes $7.55$.

This testing illustrates that it is possible to build an end-to-end model, purely based on behavioral characteristics of the attack, that can perform better than baseline guessing for the task of cyber-attack attribution. However, the data demonstrates that off-the-shelf models are not sufficient as an automated tool to apply to high-stakes settings such as attribution. However, further research is required to understand how these end-to-end workflows could augment a human's decision making process. This framework presented going from TTP identification to threat actor attribution acts as a proof-of-concept for potential decision support tooling for threat intelligence analysts and decision makers.



\begin{table}[t]
\centering
\caption{Summary of Model Performance Under Different Conditions}\label{tab:model_performance}
\begin{tabular}{lr}
\toprule
\textbf{Conditions} & \textbf{Average Rank} \\
\midrule
Baseline& 15 \\

Uniform Expert Prior & $10.68 \pm 0.53$ \\

Expert Expert Prior & $7.75 \pm 0.09$ \\

HyDE and Expert Prior & $\mathbf{7.55 \pm 0.21}$ \\

\bottomrule
\end{tabular}
\end{table}










\section{Discussion}

Analyzing the model's performance on attribution across various threat groups, we can see patterns in the threat groups who performed better or worse on attribution tasks. These trends help illustrate the qualities of a group that make it more ``attributable'' in the broader machine-learning context.

More specifically, the two groups that performed poorly in attribution were \textit{Lazarus Group} and \textit{menuPass}. These threat actors also demonstrated high average cosine similarity relative to other threat actors' TTP sets, meaning that the TTPs assigned to these threat actors were similar to those that were chosen for other threat actors. They also demonstrated high average entropy, which means that the codes assigned to these threat actors were also ``all-inclusive," or contained a lot of different codes from across the TTP list, making it difficult for the VE model or the LLM method to find underlying trends.

If we exclude these two outlier threat actors from the dataset, we find that a pattern emerge within the dataset--giving the model more data results in better downstream attribution ranking. This also suggests that this type of technique-based attribution may not work as well for threat actor groups whose TTPs have greater similarities to those of other groups or have no significant unique behavioral attributes.



\subsection{TTP Frequency Analysis}


Interestingly, four out of five of the highest frequency TTPs generated by GPT-4 are also the highest frequency TTPs from the human-generated dataset. In fact, it is apparent that there is a positive correlation between the TTPs from the GPT-4 dataset and the TTPs represented in the MITRE dataset, as seen in \cref{fig:MitreVSGPT} below.


\begin{figure}
    \includegraphics[width=0.95\linewidth]{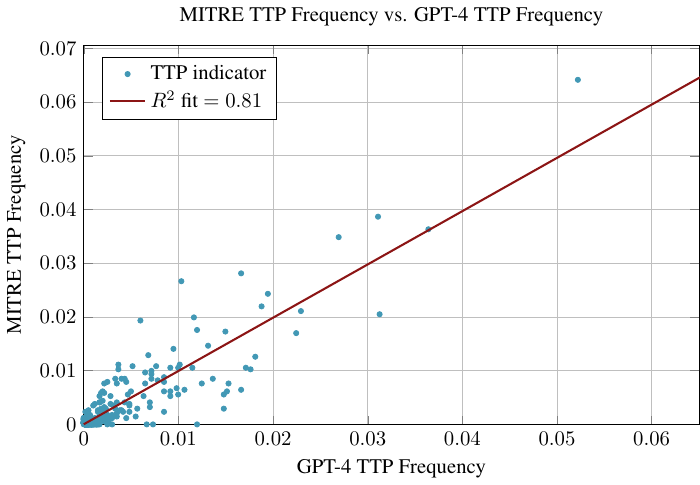}
    \caption{TTP frequencies from MITRE and GPT-4 generated datasets.}
    \label{fig:MitreVSGPT}
\end{figure}

Although the GPT-4 dataset presents a significant amount of ``noise" that may make it hard to parse for an analyst, the overall TTPs represented by the GPT-4 dataset still correlate in pattern with the MITRE dataset. This could be because GPT-4 was trained on a large corpus of data that potentially included the MITRE dataset, or that despite the additional noise that makes the Jaccard similarity low, GPT-4 is capable of generating TTPs that align with human-generated patterns, and selecting more ``general" TTPs that pair with documents.

The GPT-4 generated TTPs may be able to provide insights that can complement the MITRE database, aiding in threat actor analysis and attribution. However, from the low similarity rates seen, this tooling must be used with caution, and work to augment, rather than to replace, existing datasets like MITRE and the analyst's own interpretation of the threat actor.

\subsection{Hallucination Rates}

Overall, the rate of hallucination among the GPT-4 generated TTPs, or the amount of TTPs that did not resolve to a MITRE web-page, was also low over the $140$ threat groups. In total, GPT-4 hallucinated only $0.76\%$, or less than $1\%$ of the overall TTPs that did not exist within the MITRE taxonomy. This totaled to only $46$ cases of hallucination across all GPT-generated TTPs for $140$ threat actors. When asking GPT-4 a specific question with specific constraints for the format of the response, GPT-4 is still able to provide intelligible responses that are relevant to the field of discipline.

The hallucinations observed fell into three categories:

\begin{enumerate}
    \item Deprecated TTPs that MITRE has since merged into other techniques, likely due to the model's April 2023 training cutoff. Most frequent was ``T1064.001'', or ``Scripting," which MTIRE merged with ``T1059," ``Command and Scripting Interpreter" \cite{MITREATTCKT1064, MITREATTCKT1059}.
    \item Valid technique names paired with incorrect ID numbers. For example, with threat actor Dragonfly, GPT-4 generated ``T0854.007, Compromise Software Dependencies and Development Tools: Subversion Repository'' when the correct ID is ``T1195."
    \item Fabricated sub-techniques for existing technique IDs. For the Naikon threat actor, GPT-4 incorrectly generated both ``T1570.001, Lateral Tool Transfer'' and ``T1570, Lateral Tool Transfer'' when only the latter exists in MITRE's database without sub-techniques.
\end{enumerate}

Overall, the outputs that contain these various types of ``hallucinations" are contained within the APT actors for which there were high set differences for items ``In GPT, not MITRE". Thus, reducing the overall ``noise" of the GPT-generated reply may reduce the amount of hallucinated TTPs produced in the output.

\subsection{Comparing GPT-4 and VE Search for TTP Identification}

\begin{figure*}
    \centering
    \includegraphics[width=\textwidth]{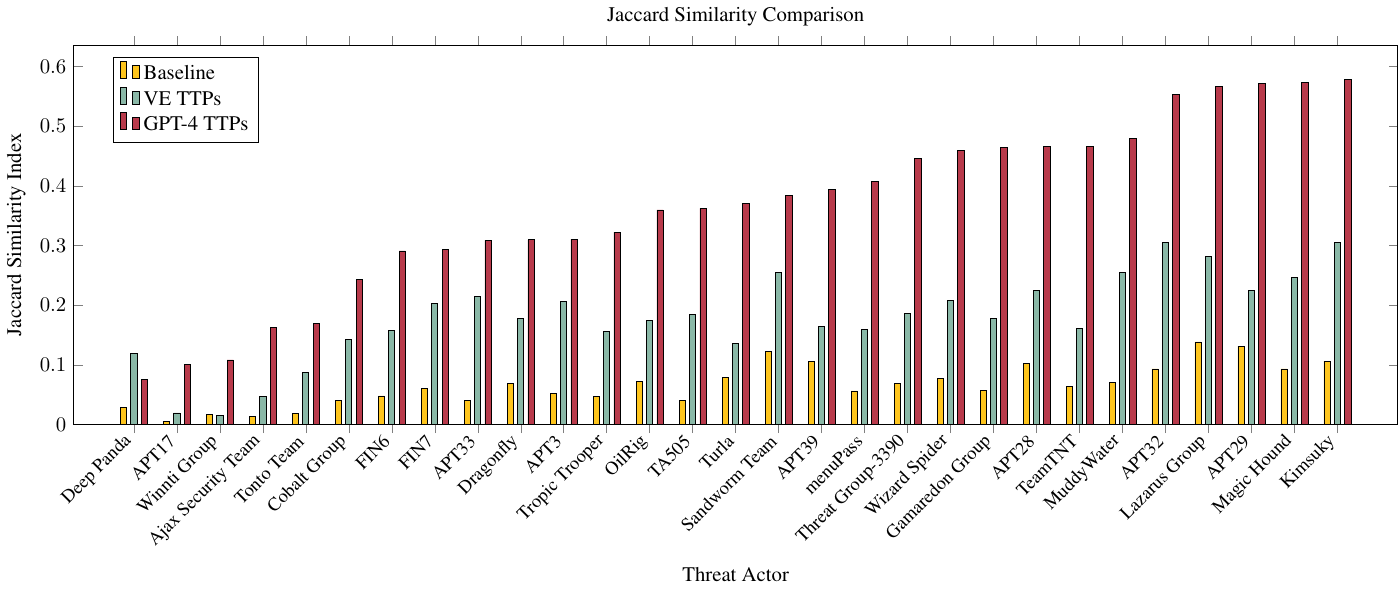}
    \caption{Comparing Jaccard similarity for the exhaustive baseline, the vector embeddings (VE), and GPT-4.}
    \label{fig:jaccard}
\end{figure*}

Shown in \cref{fig:jaccard}, Jaccard similarity was first used to calculate similarities between the GPT-4 generated TTPs and a baseline group of all TTPs from the MITRE set as a baseline scoring metric. This baseline measurement against the group of all MITRE TTPs together was known as the \textit{exhaustive baseline}, because the set that was compared to every set of GPT-4 TTPs was an exhaustive set of TTPs. This represented the worst-case scenario where GPT-4 returned every single TTP that exists within the MITRE library for all $29$ threat actors. Comparing to this baseline, both the VE and GPT approach outperform this baseline, which, on average, results in around $6.60\%$ Jaccard similarity.

Overall, GPT-4 performs with greater Jaccard similarity to the MITRE dataset for all the $29$ threat actors observed in the dataset. In the case of the VE approach, there is even more noise than the GPT-4 approach: this is likely because for the VE approach, the method simply parsed all the text files at once, without necessarily discriminating if the lines read were lines related to the title/date/author or the ending paragraph, or actual threat intelligence text.

Both approaches found around 300 different TTPs across documents, but the types of TTPs were different, with the VE TTPs having greater specificity to the attack and persistence methods. This may be because GPT-4 may have biased priors, knowing that certain TTPs are more generally used and using that to categorize a report. In other words, GPT-4 may produce more general TTPs than the TTPs produced by VE search. 

Additionally, the use of normalized weights representing the $P(\tilde{y}_\text{ttp} \mid t)$ generated by the algorithm, showcases that through the VE approach, it is possible to extract TTPs in a manner more transparent than a black box LLM output. Additionally, these outputs provide more information on the threat actor behavior than the current standard of a binary inclusion or exclusion of the TTP as a part of a threat actor's profile \cite{mitre2023techniques}. This probabilistic weight matrix also differs from the binary weights matrix presented in prior literature \cite{egyptian}.

The VE pipeline presents a powerful approach for analysts because it represents a probabilistic way of looking at TTPs through relative rankings to one another, rather than just seeing whether a threat actor uses or does not use a TTP to conduct attacks, which is what MITRE and GPT-4 produce. Additionally, in comparison with GPT-4, the VE pipeline is less of a black box and has less access to external training data that could potentially bias initial predictions for the analyst. This also demonstrates that it is possible to balance automation and transparency with accuracy. 




\section{Conclusion}

This research demonstrates that GPT-generated and VE-generated TTPs have large variance from human-generated datasets, but this does not preclude them from being useful for training a model to conduct TTP-based attribution. The best-performing model returns an average rank of $7.55$ out of $29$ total threat actors, suggesting that the model is forming relative understandings of various techniques used by threat actors through the outputs of the VE approach, and is able to generalize these learnings moderately well on unseen documentation. When looking closer at the cases where the model performed better, this study finds that giving more data to generate TTPs for a threat actor results in better downstream attribution ranking, and overall performance on a threat group relies on the overall entropy and TTP set. The model was able to use solely technical behavioral characteristics for attribution, effectively showing a proof-of-concept for reducing the problem down to finding good indicators for individual threat actor relevance. 


This study also highlights the benefits and limitations of each LLM-based approach, including the relative transparency for analysts using the VE approach as compared to the GPT-approach (via the relative rankings of TTPs as opposed to a binary approach), and broader prior context-windows when using the GPT-4 approach. Both approaches generate many TTP sets for each threat actor that are not represented in the human-generated TTP data, but the overall frequencies of each TTP generated by the GPT approach present a positive correlation with those in the MITRE dataset.

LLMs have potential to act as a decision-support tool in attribution settings, but not intended to automate decision making. This study revealed that human inputs remain a large factor in determining the accuracy of the attribution decision. Significant improvements were observed when the expert prior was introduced in both cases of using the VE-trained model for attribution, as well as GPT-4. These developed pipelines present promise in synthesizing vast amounts of documentation to provide a decision advantage to the user. LLM-based approaches may provide decision advantages by: listing unique TTPs beyond MITRE's generic ones, highlighting underlying patterns not readily visible to analysts, and adapting based on user context to offer alternative perspectives.




There are several directions for future work. One direction involves evaluating the effects of fine-tuning an off-the-shelf LLM model using human analyst TTP feedback to improve model performance on TTP identfication. Another direction is using dynamic representations of threat actor relevance, incorporating signals from open-source intelligence feeds. Additional testing may evaluate pairwise cosine similarity between VE TTPs and the GPT-4 generated TTPs as an additional test to understand the unique and differentiating nature of TTPs produced by the two approaches. 



\section*{Acknowledgment}
We would like to thank the Stanford Intelligent Systems Laboratory (SISL) and Stanford's Center of International Security and Cooperation (CISAC) for the institutional support for this project. Additionally, this testing would not be possible without inputs from Gaute Friis, Raghav Samavedam, Ankush Dhawan, the CrowdStrike team (Mark Schloesser, Tanya Widen, and Sven Krasser), and Edir Garcia Lazo--sincerest thanks for all of your guidance throughout. Thank you Dr. Cameron Tracy for your revisions on the full-length thesis version of this paper. In memory of Dr. Rodney Ewing, whose teachings during my time in the CISAC Honors program profoundly shaped my approach to research.
\printbibliography

\immediate\write18{echo "Forced compilation"}

\end{document}